\documentclass{article}

\usepackage{arxiv}

\usepackage{amsmath}
\usepackage[utf8]{inputenc} % allow utf-8 input
\usepackage[T1]{fontenc}    % use 8-bit T1 fonts
\usepackage{hyperref}       % hyperlinks
\usepackage{url}            % simple URL typesetting
\usepackage{booktabs}       % professional-quality tables
\usepackage{amsfonts}       % blackboard math symbols
\usepackage{nicefrac}       % compact symbols for 1/2, etc.
\usepackage{microtype}      % microtypography
\usepackage{lipsum}
\usepackage{graphicx}
\graphicspath{ {./images/} }

\title{Redundant Field Survey Data of Cyclotron with Imperfect Median Plane}

\author{
 Lige Zhang \\
  TRIUMF\\ 
  4004 Wesbrook Mall, Vancouver, British Columbia, V6T 2A3, Canada\\
 \texttt{lzhang@triumf.ca} \\
 \AND
Yi-Nong Rao \\
 TRIUMF\\ 
 4004 Wesbrook Mall, Vancouver, British Columbia, V6T 2A3, Canada\\
}

\begin{document}
\maketitle
\begin{abstract}
An accurate and detailed field map is important for cyclotron beam dynamics studies. During the long history of cyclotron studies, many techniques have been developed by cyclotron pioneers for the treatment of median plane field map. In this paper, we take the TRIUMF 500 MeV cyclotron as an example to study the asymmetric field resulting from the imperfect median plane symmetry. The ``Gordon approach'' and a highly accurate compact finite differentiation method are used to investigate the historical field survey data. The redundancy in the survey data is revealed by the expansion method, which also makes it possible to correct the error in the measurement. Finally, both the azimuthal field $B_\theta$ and the axial gradient of the axial field $dB_z/dz$ in the median plane are corrected using the radial field map $B_r$. The influence of the correction is examined by recalculating the equilibrium orbit properties of the TRIUMF cyclotron. The result shows significantly increased vertical centering errors of the closed orbits.  Further simulation study suggests that these centering errors can be reduced to below 1.5 cm by adjusting the trim coils' $B_r$ field within the output limits of our trim coils' power supplies. The error in the measurement field data may explain why the calculated trim coils' settings during the cyclotron commissioning in 1974 encountered difficulty.
\end{abstract}

% keywords can be removed

\section{Introduction}\label{sec:IP1}

We would like to start the introduction with some clarification on the cyclotron type~\cite{craddock2008cyclotrons, chautard2006beam} discussed in this paper. The cyclotron mentioned hereinafter refers to an isochronous cyclotron with edge focusing, which is one of the most widely used types of fixed field accelerators nowadays.  

Beam dynamics study of the cyclotron requires a very accurate field map of large complex main magnet~\cite{chautard2006beam, kleeven2015cyclotrons}. This differs from the beam dynamics modelling in synchrotrons; in synchrotrons, the reference orbit is fixed and pre-defined; having a periodic lattice mainly consisting of dipoles and quadrupoles, in the first approximation it consists only of straight lines and circular arc segments. In cyclotrons, this simplification almost never applies; neither the magnetic field nor its radial gradient (index) are constant over any given segment of the closed orbit. The field is global, varying with radius and azimuth and so closed orbits can only be found by integrating the equations of motion. Moreover, the isochronism condition should be met to an accuracy of one part in $10^4$ to $10^5$ over its complete momentum range of a factor 30 or more, whereas in synchrotrons there is no such constraint and the aperture needs to accommodate a momentum range of only a few percent. Thus, a feasible way to study the beam dynamics in the cyclotron is by integrating the motion equations through a detailed and accurate field map.

An accurate 2D or 3D field map of the cyclotron main magnet can be obtained from measurement or finite element analysis (FEA) software computation. Usually, a field expansion from the 2D median plane field~\cite{Hart:2011rq} is sufficient for the beam dynamics study, as the beam in the cyclotron spirals out radially with the vertical position constantly oscillating around the median plane, and the beam's vertical extent is several times smaller than the magnet gap. Different methods of field expansion out of a 2D plane have been used in particle tracking codes. For example, {\tt Zgoubi}~\cite{meot2012zgoubi} uses Taylor expansion with curl and divergence of the magnetic field being zero in cartesian coordinates. {\tt CYCLOPS}~\cite{Gordon:1984EO} and {\tt Goblin}~\cite{gordon1959ornl} use the expression in cylindrical coordinates, which is derived from a scalar potential satisfying Laplace's equation. This method was found in Mort Gordon's notes and had been organized in Baartman's cyclotron book~\cite{Rick2021}. The Gordon approach gives a clear physical insight into the median plane symmetric field and the asymmetric field resulting from a tilted median plane. Using this expression, we find the redundancy in the survey data of TRIUMF 500 MeV cyclotron median plane asymmetric field.

To avoid non-physical artefacts in the integration of motion equations, the numerical expansion of the median plane field should satisfy Maxwell's equations to sufficient order. For the field expansion that only includes the first-order of the vertical displacement, the derivative calculated from the  interpolation polynomial is accurate enough for the integrator~\cite {TRI-DN-70-45}.  For the higher-order expansion, some codes such as {\tt COSY}-$\infty$~\cite{makino2006cosy} interpolate the field in the radial direction by using a gaussian basis function, the derivative thereby can be calculated to any order analytically. But on the other hand, the resulting errors in the derivatives depend sensitively on the parameters of the gaussian basis function, and the high order derivatives contain magnified high-frequency noise; the errors can hardly be reduced. 

Some other codes use numerical differentiation to calculate high-order derivatives, in which the errors of the derivatives are bound on the points of the original sampling data. The differentiator can also be generalized as an interpolation function to calculate the given order derivative between the sampling points. The $n^{\rm th}$ order derivative interpolation function is generated from the $n^{\rm th}$ order differentiator instead of the derivative of the $(n-1)^{\rm th}$ order interpolation function. Thus the numerical differentiation has more degrees of freedom to reduce the errors of the high order derivatives than the conventional interpolation method. A detailed study for the numerical differentiator was undertaken by Dong-o Jeon~\cite{JEON1997167}. This method successfully improved the numerical accuracy: for example the {\tt Z$^4$} orbit code that employs Jeon's differentiator can reduce the symplectic error in simulating the high order resonance passage~\cite{gordon1986z4}. In this paper we'll be using Jeon's compact finite differentiator to correct the field survey data.

In the TRIUMF 500 MeV cyclotron, the field survey was taken in the geometrical median plane on a polar grid using a flip coil mapping system ~\cite{auld1972magnetic}, consisting of 105 equally-spaced coils mounted on a 8.2\,m aluminum I-beam oriented radially and pivoting in one degree increments at the cyclotron centre. The data of the axial field, its axial gradient, the radial and azimuthal field components were taken. For a cyclotron with perfect median plane symmetry, the radial and azimuthal field components as well as the axial gradient of the axial field component in the median plane all are zero. But because of the pole's geometric error and the unevenly magnetized soft iron, the magnetic median plane where there are zero transverse field components irregularly deviates by a few cm from the geometric median plane, giving rise to non-zero asymmetrical fields in the geometric median plane. The field survey data that we inherited consists of two files; one has the median plane axial field which was well shaped to provide proper isochronism and tunes, the other includes 3 asymmetric fields, $B_r$, $B_\theta$ and $dB_z/dz$, which were directly interpolated in {\tt CYCLOPS} for integrating the motion equations. These field data are given as Fourier series in azimuth truncated at a given order to reduce the high frequency noise. We do not have the raw data from the field survey but only the processed data. Using Gordon's field expansion technique~\cite{gordon1986z4}, we have discovered self-consistency errors in these field data. However, these are easily and convincingly corrected, as will be shown.

The median plane asymmetric field in a small gap cyclotron or cyclotron with a high magnetic field is usually negligible, and in that case, the surveyed or calculated axial field $B_z$ in the median plane is usually sufficient for the beam dynamics study. But for a large gap magnet with low magnetic field, such as the TRIUMF 500 MeV cyclotron, the asymmetric field can significantly shift the vertical position of the beam. Moreover, the error of the tilted median plane can be the driving force when the tunes pass through coupling resonances~\cite{gordon1962effects}. To study the effect of the asymmetric field on the beam dynamics, only the field survey data can be used while the finite element analysis calculation can not reveal the pole machining errors and the variations of material properties of the steel. (The steel was sourced from the Davie ship-building company of Quebec.) The tolerance of the error field was carefully studied in 1972~\cite{bolduc1972}, and the cyclotron was commissioned in 1974-1975~\cite{Richardson1975,craddock1975properties}, though the passage through coupling resonances has not been studied until recently. After correcting the errors in the measurement data, we have recalculated the properties of static equilibrium orbits. Further, we have optimized the trim coils' settings to achieve a better beam vertical centering.

%Since the median plane error field is important for the beam dynamic study of the TRIUMF 500 MeV cyclotron. In this paper, we investigate the survey data of the TRIUMF cyclotron. The redundancy in the survey data is revealed by using field expansion techniques, which is also used to check and correct the error in the field measurement data. Simulation study of the static equilibrium orbit properties is conducted using the corrected field map. Further in the simulation study, we optimized the trim coil setting to achieve a better beam vertical centering.

\section {Field expansion out of median plane: Gordon's approach}\label{sec:IP2}

The magnetic field map must satisfy Maxwell's equations to sufficient order. In a cyclotron, a typical way is to expand the field relative to the median plane. Gordon's approach is one of these; it is derived from the scalar potential $\Psi$ that satisfies Laplace's equation. By solving the 3D Laplace equation using an operator trick, we get the potential $\Psi$ expanded in powers of axial position $z$ as follows~\cite{Rick2021}
\begin{equation}
\begin{aligned}   
        \Psi&=\Psi_{\rm o}+\Psi_{\rm e} ,\\
        \Psi_{\rm o}&=zB-\frac{z^3}{3!}\nabla_2^2B+\frac{z^5}{5!}\nabla_2^4B-... ,\\
        \Psi_{\rm e}&=C-\frac{z^2}{2!}\nabla_2^2C+\frac{z^4}{4!}\nabla_2^4C-...,
\end{aligned}
\end{equation} where $B=B(r,\theta)$ and $C=C(r,\theta)$, and where $\nabla_2^2$ is the 2-dimensional Laplace operator 
\begin{equation}\label{eq:2d}
\nabla_2^2\Psi=\frac{1}{r}\frac{\partial}{\partial r}\left(r\frac{\partial\Psi}{\partial r}\right)+\frac{1}{r^2}\frac{\partial^2\Psi}{\partial\theta^2}. \end{equation} Thus, for example, $\nabla^2\Psi=\frac{\partial^2\Psi}{\partial z^2}+\nabla_2^2\Psi=0$.

The odd term $\Psi_{\rm o}$ produces a field with median plane symmetry, while the even term $\Psi_{\rm e}$ spoils this symmetry. The magnetic field is given by $\vec{B}=-\nabla\Psi$, that is 
% Since the part of the magnetic field generated by $C$ corresponds to the median plane error, a field with perfect median plane symmetry could be given by taking $C=0$. Thus, the field off the median plane can be expressed in terms of $B$ and its derivatives, that is
% \begin{eqnarray}\label{bzex}
% \begin{aligned} 
% B_z&=-B+\frac{z^2}{2!}\nabla_2^2B-\frac{z^4}{4!}\nabla_2^4B+...\\
% B_r&=-z\frac{\partial B}{\partial r}+\frac{z^3}{3!}\frac{\partial\nabla_2^2B}{\partial r}-...\\
% rB_\theta&=-z\frac{\partial B}{\partial\theta}+\frac{z^3}{3!}\frac{\partial\nabla_2^2B}{\partial\theta}-...
% \end{aligned}
% \end{eqnarray}
%
% If we only consider the field generated by $C$, by taking $B=0$ the median plane error field could be given as
% \begin{eqnarray}\label{brex}
% \begin{aligned} 
% B_z&=z\nabla_2^2C-\frac{z^3}{3!}\nabla_2^4C+...\\
% B_r&=-\frac{\partial C}{\partial r}+\frac{z^2}{2!}\frac{\partial\nabla_2^2C}{\partial r}-...\\
% rB_\theta&=-\frac{\partial C}{\partial\theta}+\frac{z^2}{2!}\frac{\partial\nabla_2^2C}{\partial\theta}-...
% \end{aligned}
% \end{eqnarray}
\begin{eqnarray}\label{bzex}
\begin{aligned} 
B_z&=-B+z\nabla_2^2C+\frac{z^2}{2!}\nabla_2^2B-\frac{z^3}{3!}\nabla_2^4C-\frac{z^4}{4!}\nabla_2^4B+... ,\\
B_r&=-\frac{\partial C}{\partial r}-z\frac{\partial B}{\partial r}+\frac{z^2}{2!}\frac{\partial\nabla_2^2C}{\partial r}+\frac{z^3}{3!}\frac{\partial\nabla_2^2B}{\partial r}-... ,\\
rB_\theta&=-\frac{\partial C}{\partial\theta}-z\frac{\partial B}{\partial\theta}+\frac{z^2}{2!}\frac{\partial\nabla_2^2C}{\partial\theta}+\frac{z^3}{3!}\frac{\partial\nabla_2^2B}{\partial\theta}-...  .
\end{aligned}
\end{eqnarray}
In most orbit programs, $C$ is ignored and only the zero-order $B_z$ value and the first-order $B_r$ and $B_\theta$ values are used. This is acceptable only for $z$ very small compared with the magnet gap since it violates $\nabla\cdot\vec{B}=0$ and can therefore lead to non-physical results for finite $z$ values. This can be remedied by including the $z^2$ term in $B_z$. In general, when $B_r$ and $B_\theta$ are given to order $z^n$, then $B_z$ should be given to order $z^{n+1}$.

Ignoring $C$ is appropriate in the initial design stage of a cyclotron, but not for for finding tolerances for manufacturing errors, nor for detailed investigations of orbit excursions and resonance crossings in an as-built cyclotron. In the TRIUMF cyclotron, the vertical closed orbit excursion is as large as $\pm1.3$\,cm even after correction, as shown below. In a synchrotron, the closed orbit distortion is corrected with separate small dipole magnets. But in cyclotrons, where the radial extent of the beam gap is orders of magnitude larger than that gap, this is not possible; instead, orbits are corrected vertically using trim coils that are placed above and below the median plane. Powered in opposition to each other, they create radial field components that can correct the beam vertical position.
  
\section{Numerical differentiation} \label{sec:IP3}

To calculate the coefficients of the expansion series in eq.\,~\eqref{bzex}, numerical differentiation algorithms are needed to evaluate the derivatives of the median plane field. Initially, a simple second-order central difference scheme was used~\cite{gordon1986z4},
\begin{eqnarray} \label{cd}
        \begin{aligned}
         f_{i}^{\prime}&=\left(f_{i+1}-f_{i-1}\right) / 2 d ,\\
                f_{i}^{\prime \prime}&=\left(f_{i+2}+f_{i-2}-2 f_{i}\right) / 4 d^{2} ,
        \end{aligned}
\end{eqnarray} 
where $d$ is the distance between adjacent data points in the field map. Afterwards, a so-called `compact finite difference method' (CFD) was developed to improve the accuracy of the calculated derivatives, and was implemented in the Z$^4$ cyclotron code~\cite{jeon1995thesis,JEON1997167}. Here we directly use this technique and the formulas. Suppose that there is a uniform mesh onto which a sinusoidal signal $e^{jkx}$ is applied, with $k$ denoting the wave-number. The phase advance between consecutive two points is $kd$. We define $\omega:=kd$. The largest $\omega$ without the problem of "aliasing" is $\pi$. The frequency response of the low-pass filter employed to reduce the high-frequency noise in the measurement data is represented as 
\begin{equation}\label{lf}
        H^{(F)}(\omega)=\frac{a+b \cos \omega+c \cos 2 \omega+d \cos 3 \omega}{1+2 \alpha \cos \omega+2 \beta \cos 2 \omega}.
\end{equation}
By imposing the 6$^{\rm th}$ order accuracy constraint, the coefficients in the low-pass filter are given as \begin{equation}
        \begin{aligned}
        &\alpha=0, \quad \beta=\frac{3}{10}, \quad a=\frac{1}{2} \text {, }\\
        &b=\frac{3}{4}, \quad c=\frac{3}{10}, \quad d=\frac{1}{20}.
        \end{aligned}
\end{equation}
%For the given filter and the first order differentiator, it is formally sixth-order accurate. 
The transfer function of the first order differentiator is written as  \begin{equation}\label{cfd1}
        H^{(1)}(\omega)=j \frac{(14 / 9) \sin \omega+(1 / 18) \sin 2 \omega}{1+(2 / 3) \cos \omega},
\end{equation}
while the transfer function of the second order differentiator (accurate to fourth order) is   
\begin{equation}\label{cfd2}
        H^{(2)}(\omega)=\frac{(24 / 11)(\cos \omega-1)+(3 / 22)(\cos 2 \omega-1)}{1+(4 / 11) \cos \omega}.
\end{equation}

\begin{figure}
        \centering
        \includegraphics[width=1.\textwidth]{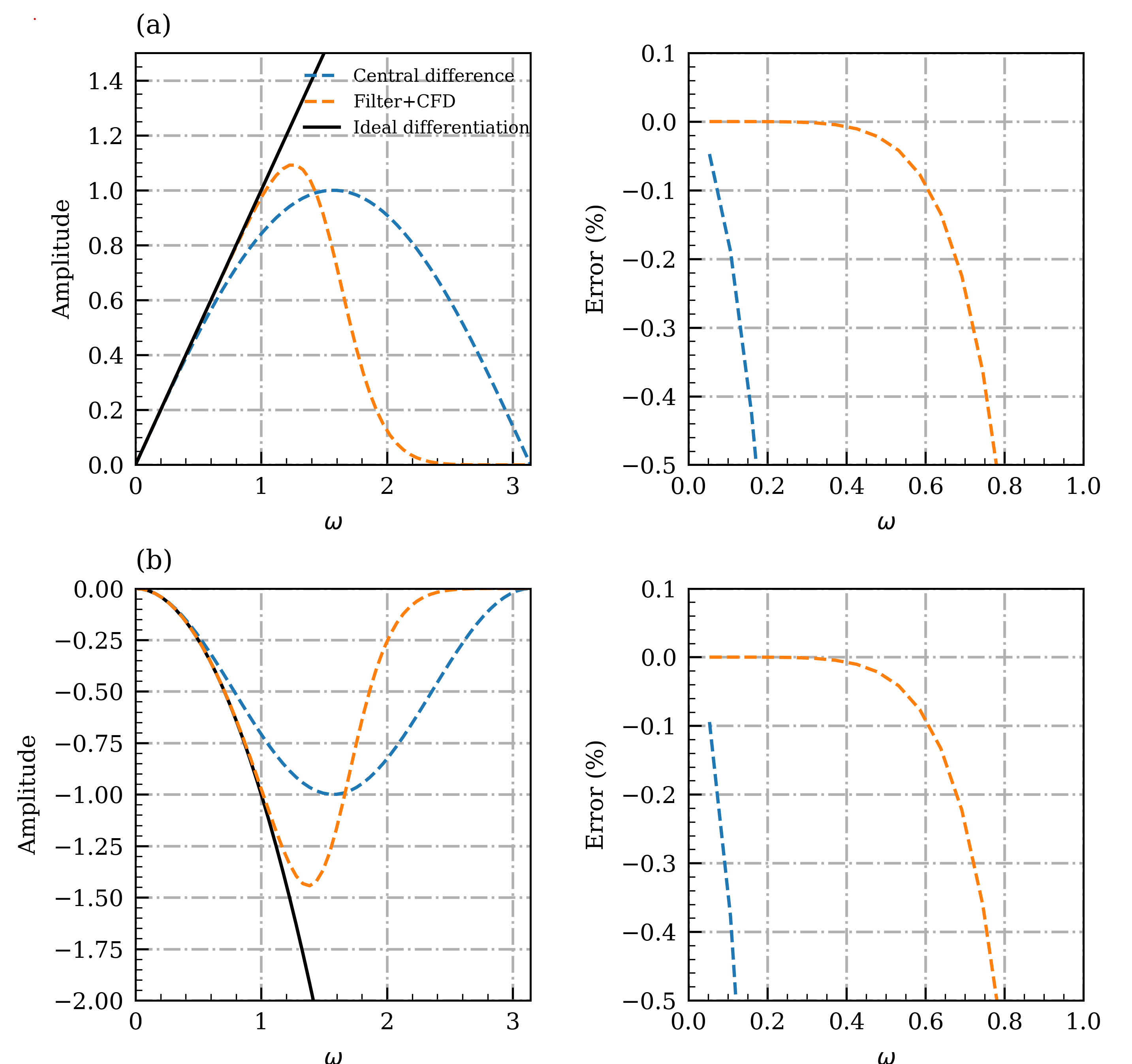}
        \caption{(a) Frequency response and error of the first order differentiator. (b) Frequency response and error of the second order differentiator. The ideal frequency response of the differentiator is given by the mathematical transfer function of derivation. The blue dashed line is the result of the central difference scheme given by Eq.\,\eqref{cd}, while the orange dashed line is the result of the compact difference differentiator.}\label{fr}
        \end{figure}
        
The product of the frequency response of the low-pass filter and that of the intermediate first or second order differentiator gives the total frequency response of that differentiator. Fig.\,\ref{fr} compares the frequency response and error of the differentiators. The compact finite difference differentiator gives zero error at low frequency. It also has a sharp cut-off at high frequency. For the TRIUMF cyclotron, the amplitude of the asymmetric field from the tilted median plane is small. The final survey data was given as coefficients of Fourier series in azimuthal direction containing up to $29^{\rm th}$ harmonic~\cite{TRI-DN-70-45}. The error of the differentiators is less than 0.05\%  for the harmonics smaller than the $29^{\rm th}$ ($\omega =:k d=0.51$, $d=2\pi/360$). Thus we can expect the differentiator to handle the survey data properly.

\section{Redundant field survey data of TRIUMF 500 MeV cyclotron} \label{sec:IP4}

\begin{figure}
        \centering
        \includegraphics[width=.75\textwidth]{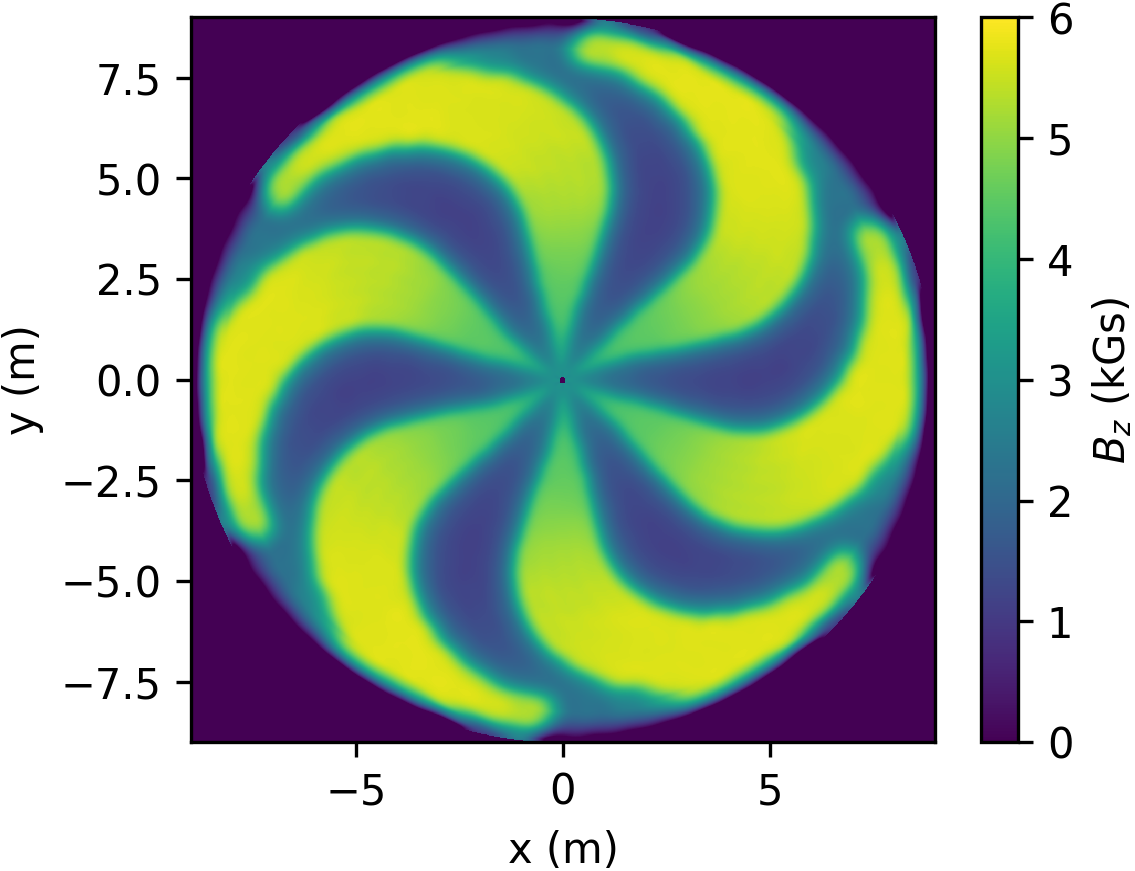}
        \caption{Axial field in the median plane of the TRIUMF 500 MeV cyclotron.}\label{BZMAP}
        \end{figure}

The symmetrical part of the cyclotron field is directly given by the measured axial field in the median plane as shown in Fig.\,\ref{BZMAP}, while the axial derivative of this field and the transverse components in the median plane all are derived from the function $C(r,\theta)$. We use this fact to correct errors in the measured asymmetric field components.
%\clearpage

\subsection{Crosschecking the redundant field survey data}
By substituting $z=0$ in eq.\,\eqref{bzex}, the magnetic field in the median plane is written as 
\begin{equation}\label{b0}
\begin{aligned}  
        B_r&=-\frac{\partial C}{\partial r} ,\\
        rB_\theta&=-\frac{\partial C}{\partial \theta} ,\\
        B_z&=-B .
\end{aligned}
\end{equation}
When the median plane symmetry is broken, the axial derivative of the axial field $dB_z/dz$ in the median plane is non-zero and is expressed as 
\begin{equation}\label{dbdz0}
\begin{aligned}
        \frac{d B_z}{d z}=\nabla_2^2C=\frac{\partial^2 C}{\partial r^2}+\frac{1}{r}\frac{\partial C}{\partial r}+\frac{1}{r^2}\frac{ \partial^2 C}{\partial \theta^2}. 
\end{aligned}
\end{equation}

\begin{figure}
        \centering
        \includegraphics[width=.85\textwidth]{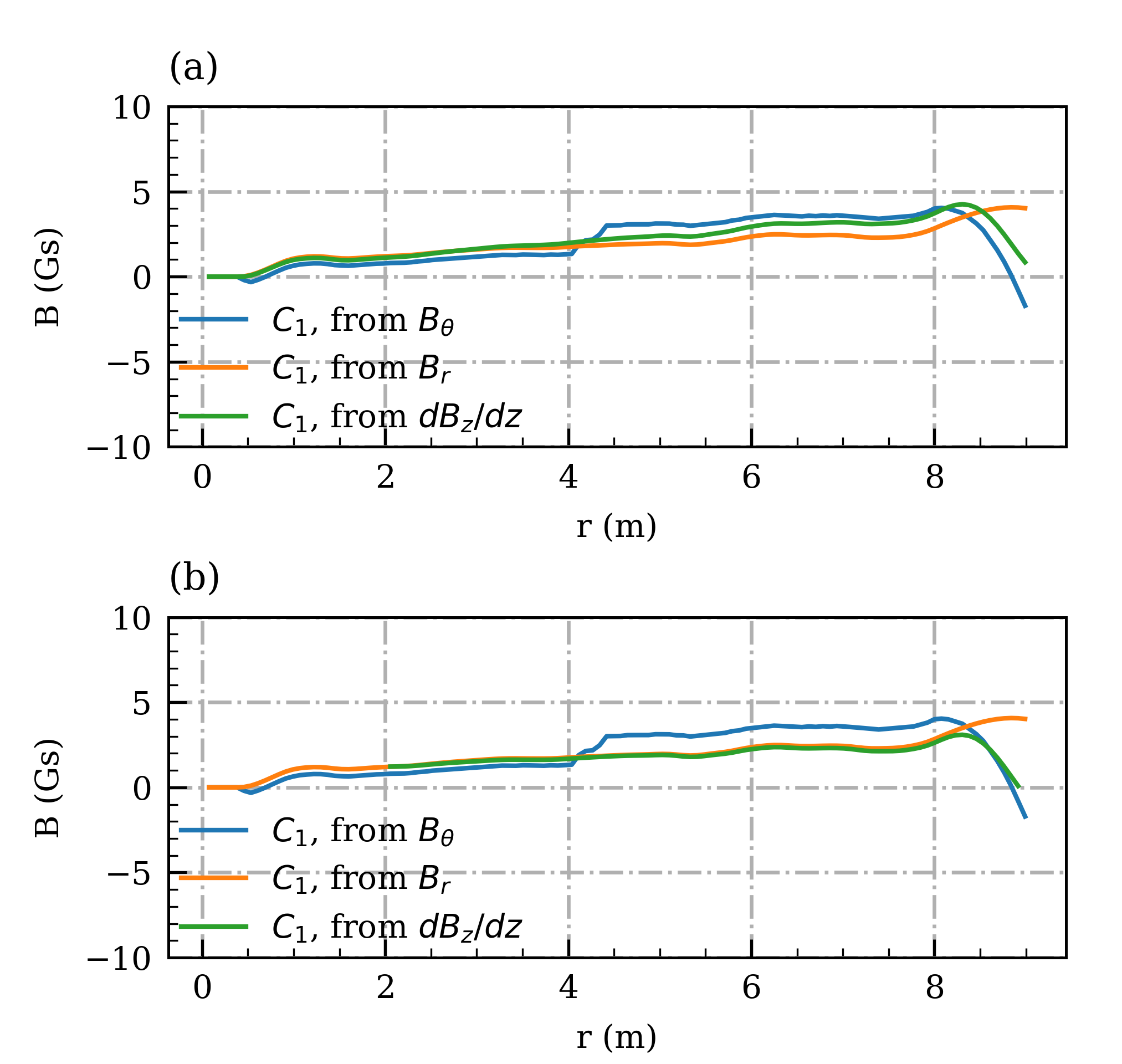}
        \caption{First harmonic of the asymmetric field potential map $C_1$. (a) Calculated using eq.\,\eqref{CHar} starting from the center of the cyclotron, where the field is homogeneous and thus the ODE's initial conditions are $C_1(0)=-jB_{\theta}(0)=0$ and $C_1'(0)=-B_{r}(0)=0$. (b) Calculated using eq.\,\eqref{CHar} starting from the radius of 2 m in the third ODE.}\label{C1}
        \end{figure}

The asymmetric components $B_r$, $B_\theta$ and $dB_z/dz$ were measured at $z=0$ on a polar grid with $1^\circ$ intervals in azimuth and 3 inch intervals in radius, but the final data are given as Fourier series in azimuth, up to 29 harmonics. Since different harmonics are orthogonal, every harmonic individually satisfies the eqs.\,\eqref{b0} and \eqref{dbdz0}. The $n^{\rm th}$ harmonic of the map $B_{rn}$, $B_{\theta n}$ and $d B_{z n}/d z$ satisfies
\begin{equation}\label{CHarO}
\begin{aligned}
        B_{rn} &=-\frac{\partial C_n}{\partial r} ,\\
        r B_{\theta n} &=-\frac{\partial C_n}{\partial \theta}, \\
        \frac{d B_{z n}}{d z} &=\frac{\partial^2 C_n}{\partial r^2}+\frac{1}{r}\frac{\partial C_n}{\partial r}+\frac{1}{r^2}\frac{\partial^2 C_n}{\partial \theta^2},
\end{aligned}
\end{equation} where $C_n$ is the $n^{\rm th}$ harmonic of the scalar potential of the asymmetric field. Writing the harmonic in complex form, eqs.\,\eqref{CHarO} are simplified to ordinary differential equations (ODE) which have only the radius $r$ as variable
\begin{equation}\label{CHar}
\begin{aligned}
        B_{rn} &=-\frac{d C_n}{d r} ,\\
        B_{\theta n} &=-jn\frac{C_n}{r} , \\
        \frac{d B_{z n}}{d z} &=\frac{d^2 C_n}{d r^2}+\frac{1}{r}\frac{d C_n}{d r}-n^2\frac{C_n}{r^2} .
\end{aligned}
\end{equation}
In the complex form of the $n^{\rm th}$ harmonic, $n$ could be either sign. So eq.\,\eqref{CHar} should be solved from $-29^{\rm th}$ to $29^{\rm th}$ harmonics.  By solving eq.\,\eqref{CHar} numerically, we get three versions of the $C$ map, from the survey data of $B_r$, $B_\theta$ and $dB_z/dz$ respectively. As an example, Fig.\,\ref{C1} compares the obtained first harmonic of the $C$ map. From $B_\theta$, there are two regions of first derivative discontinuities, occurring at $\sim$0.5 and 4 m respectively. This suggests an existence of some systematic error in the measurement data of $B_\theta$. The difference between the $C_1$ as calculated from $B_r$ and $dB_z/dz$ grows with the radius, but this is easily corrected by changing the initial condition within uncertainty. As a result of finite size of the flip coils, the survey data has a relatively larger error at smaller radius. Thus, if integrating from a larger radius with a relatively constant slope in the field, the difference becomes smaller, as shown in Fig.\,\ref{C1}(b). 

\subsection{Correcting the error in the field survey data}
The redundancy in the survey data makes it possible to correct the error in the measurement data. Using eq.\,\eqref{CHar}, we can generate a full map from a single potential map $C$. To reduce numerical errors due to the interpolation while solving eq.\,\eqref{CHar}, we directly use the CFD method to calculate the new maps. The equations used for the correction are obtained by substituting $C_n$ with $j r B_{\theta n}/n$ in eq.\,\eqref{CHar}
\begin{equation}\label{CHar1}
\begin{aligned}
        jn B_{rn} &= \frac{d (r B_{\theta n})}{d r},\\
        j n\frac{d B_{z n}}{d z} &=-\frac{d^2(r B_{\theta n}) }{d r^2}-\frac{1}{r}\frac{d (r B_{\theta n})}{d r}+n^2\frac{r B_{\theta n}}{r^2} .
\end{aligned}
\end{equation}
The numerical scheme used to solve the eq.\,\eqref{CHar1} can be easily derived using the transfer function given in eqs.\,\eqref{cfd1} and \eqref{cfd2}, and it's represented as

\begin{equation}
\begin{aligned}
\frac{1}{3} f_{i-1}^{\prime}+f_{i}^{\prime}+\frac{1}{3} f_{i+1}^{\prime}&=\frac{14}{9} \frac{f_{i+1}-f_{i-1}}{2 d}+\frac{1}{9} \frac{f_{i+2}-f_{i-2}}{4 d} ,\\
\frac{2}{11}f_{i-1}^{\prime \prime}+f_{i}^{\prime \prime}+\frac{2}{11} f_{i+1}^{\prime \prime}&= \frac{12}{11} \frac{f_{i+1}+f_{i-1}-2 f_{i}}{d^{2}}+\frac{3}{11} \frac{f_{i+2}+f_{i-2}-2 f_{i}}{4 d^{2}} .
\end{aligned}
\end{equation}

\begin{figure}
        \centering
        \includegraphics[width=.85\textwidth]{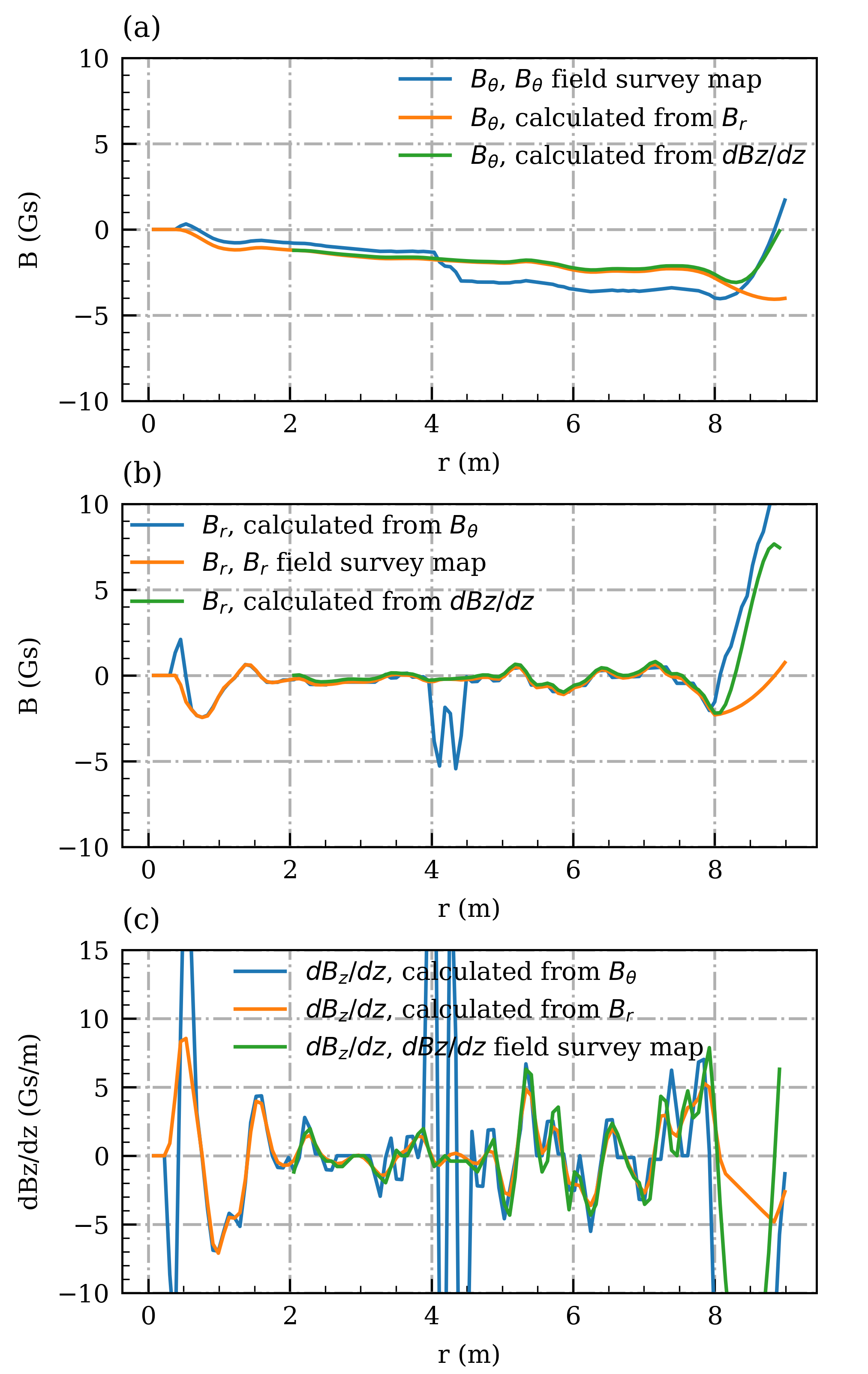}
        \caption{Comparison of the first harmonic of the asymmetric field components $B_\theta$ (a), $B_r$ (b) and $dB_z/dz$ (c), constructed with different survey data. }\label{B1}
        \end{figure}

The resulting first harmonic field is compared in Fig.\,\ref{B1}.  The $B_\theta$ survey data seems to be shifted upward at radii before 4 m and thereafter shifted downward, in comparison with those calculated from the other two survey maps. The $B_r$ survey map agrees with the calculated ones except for the spike at $\sim$4 m occurring in the one calculated from $B_{\theta}$ (see fig.\,\ref{B1} (b)). $B_\theta$ is too noisy to give a usable $dB_z/dz$ map, while the calculated $dB_z/dz$ from $B_r$ agrees with the survey result (see Fig.\,\ref{B1} (c)) and is even smoother than the survey data because high-frequency components are filtered out. 

\begin{figure}
        \centering
        \includegraphics[width=.6\textwidth]{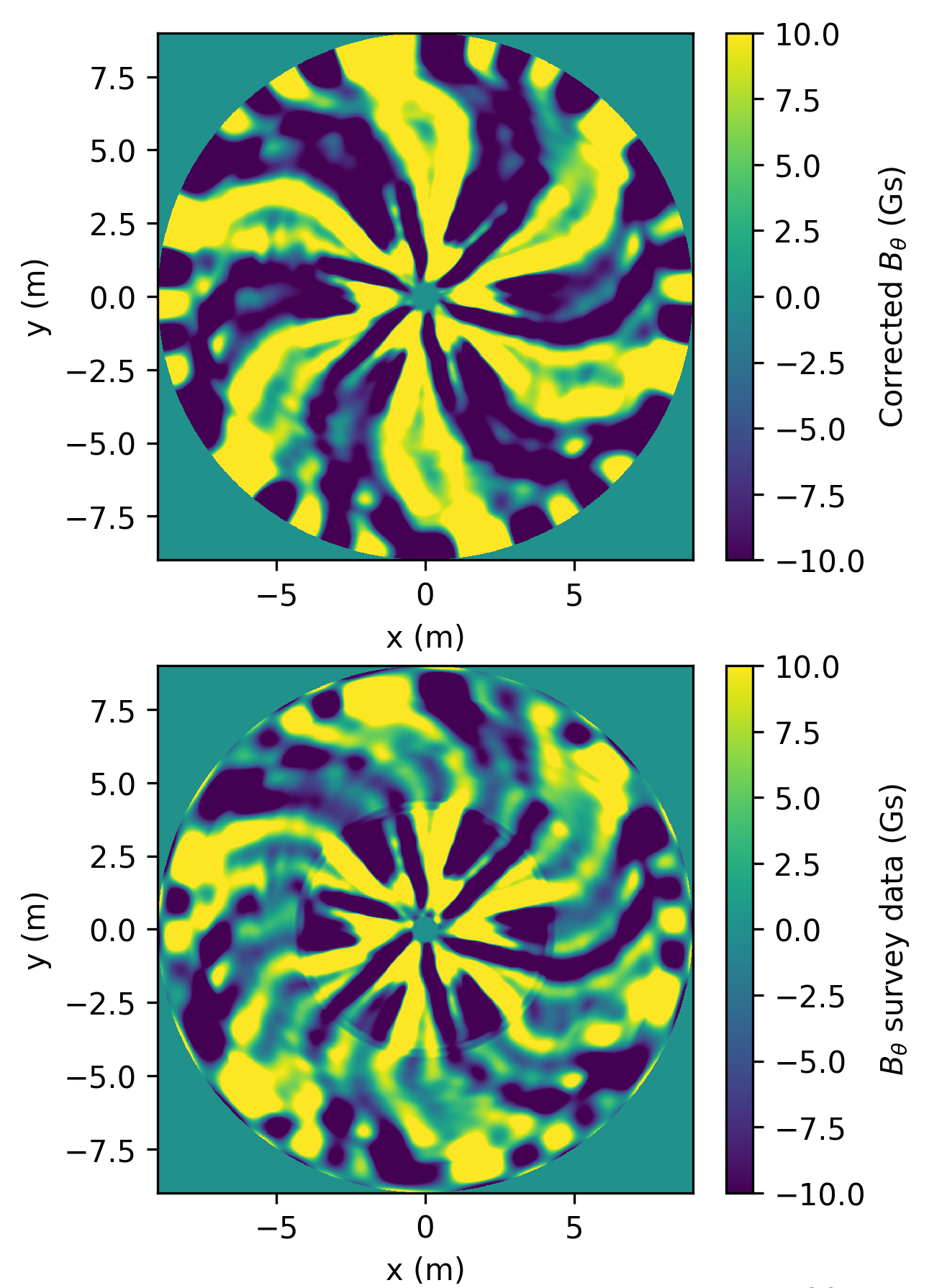}
        \caption{Azimuthal field map in the median plane. The survey map (lower) displays an obvious discontinuity at radius of $\sim$4 m, and thereafter a blurred edge of the sector structure. The corrected map (upper) shows a sharper image of the sector edges of main magnet. }\label{Bthe}
        \end{figure}

\begin{figure}
        \centering
        \includegraphics[width=.6\textwidth]{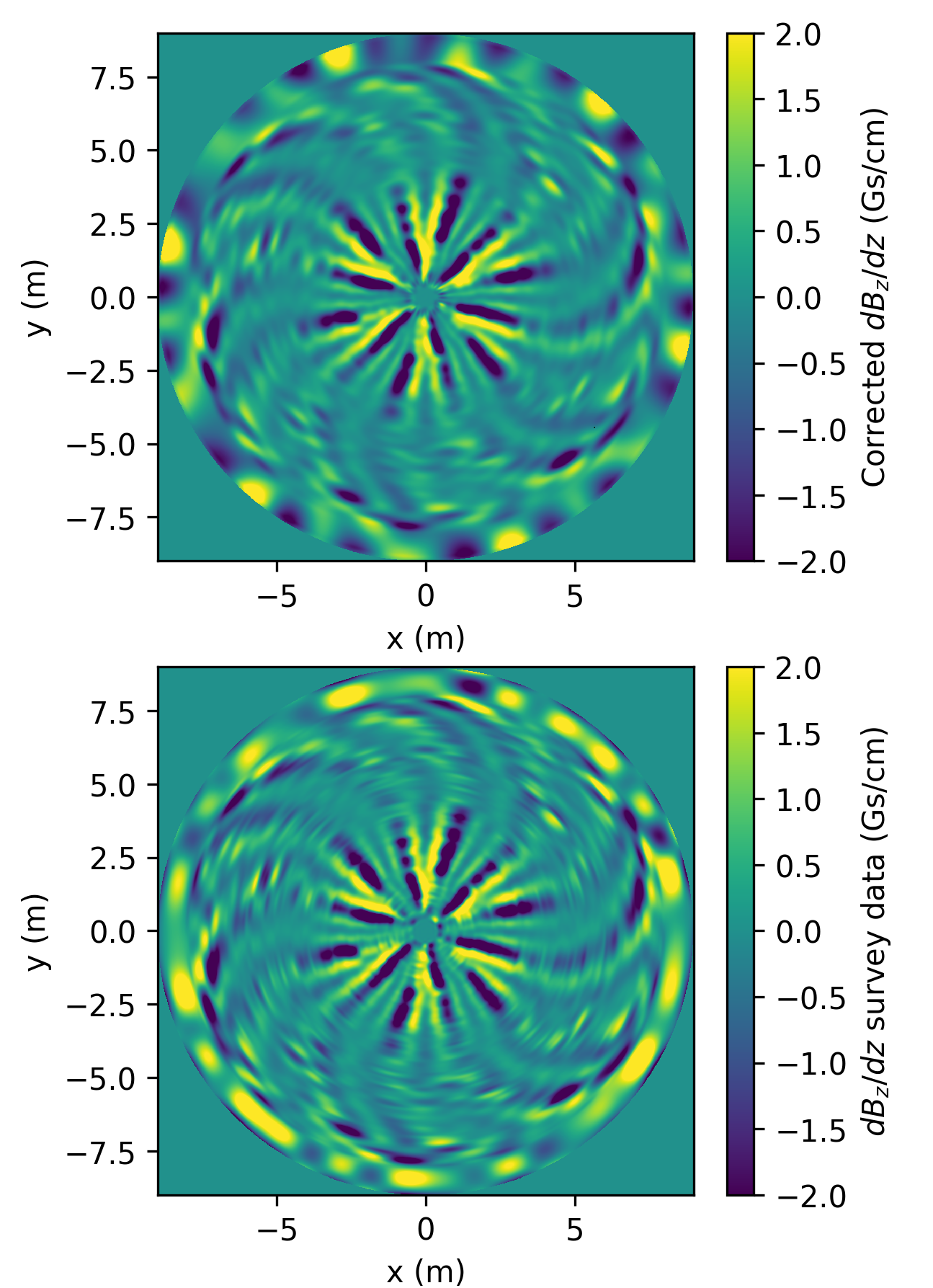}
        \caption{Axial gradient of the axial field in the median plane. The corrected result (upper) and the survey result (lower) have the same pattern where the field value changes rapidly in azimuthal direction at the edge of the pole sectors. The rms of the difference between the 2 maps in a radial range up to 8 m is 0.0013 Gs/cm, which is within the measurement uncertainty.}\label{DBDZ}
        \end{figure}

Since the magnet gap is so large at $\sim$0.5\,m, it is inconceivable that the radial field can step discontinuously on a scale of the flip coil separation ($<0.08$\,m). We therefore choose the radial field $B_r$ survey data to correct the error of the azimuthal field $B_\theta$; this is also because the reconstruction is less sensitive to the initial conditions than using $dB_z/dz$ as shown in Fig.\,\ref{C1}. Fig.\,\ref{Bthe} compares the corrected $B_\theta$ map with the survey result. To make the field satisfy Maxwell's equation with high accuracy, we also reconstruct the axial gradient of the axial field using the same $B_r$ survey data. Fig.\,\ref{DBDZ} compares the results.

\section{Static equilibrium orbit properties }\label{sec:IP5}

\begin{figure}
        \centering
        \includegraphics[width=.75\textwidth]{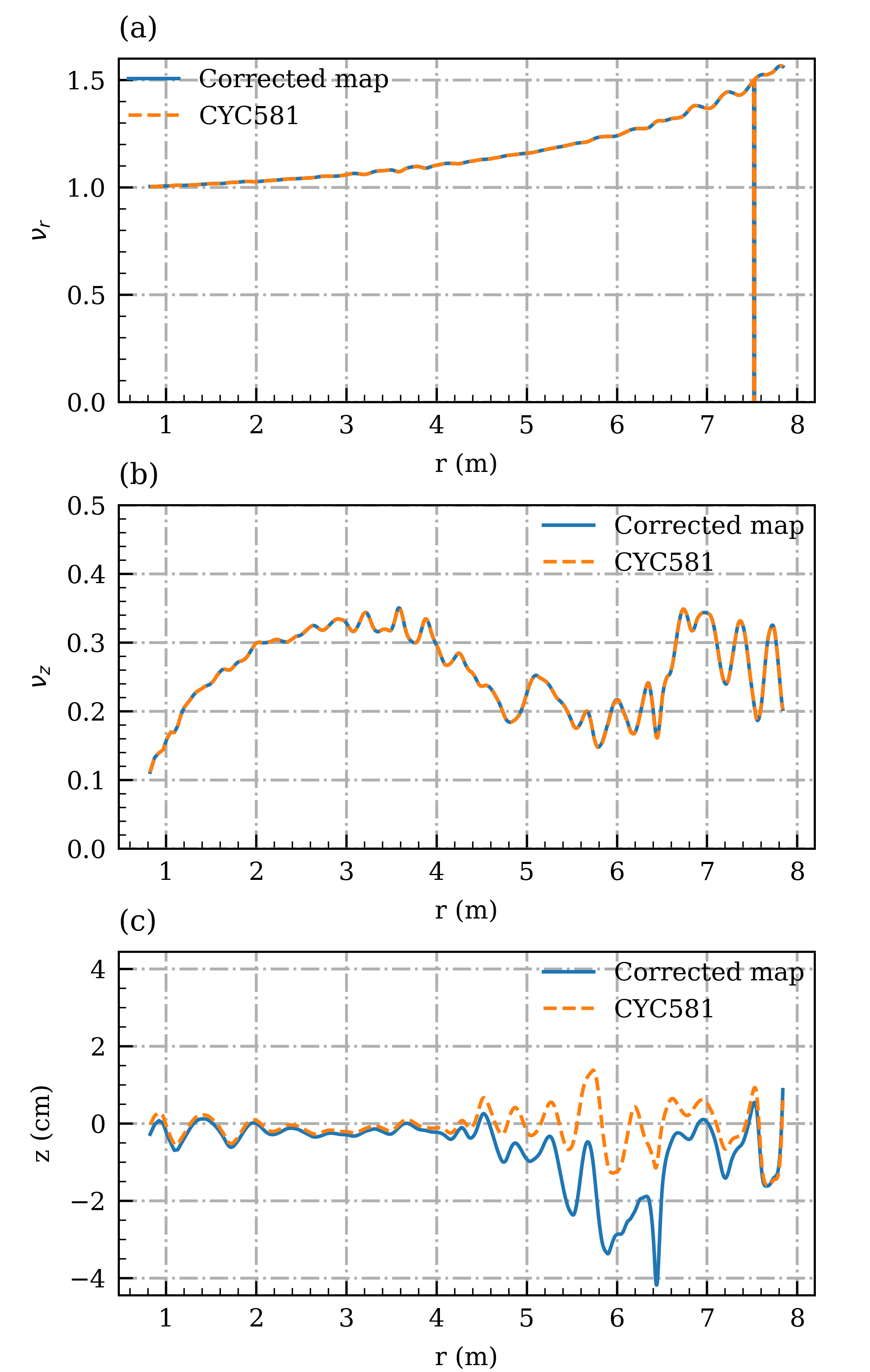}
        \caption{Tune and average vertical position of the closed orbit before and after the field error correction. The spike at the radius of $\sim$7.5 m in the $\nu_r$ is caused by the passage through the $2\nu_r=3$ resonance as on the resonance the stable region in the radial phase space is reduced to zero.}\label{CYCLOPS}
        \end{figure}

Preparatory to the commissioning work in 1974, trim coil settings were determined to correct the vertical closed orbit. These became part of the standard field survey file. Indeed, the original field files used in the closed orbit calculation find it to be corrected to $\pm 1.5$\,cm. See Fig.\,\ref{CYCLOPS}(c). This standard orbit code run became known as CYC581, and has been used as a reference for the orbit characterization through the past 48 years.

The trim coil settings predicted from the final magnet survey of April 16, 1974 turned out to be of limited value once the commissioning started in November 1974. New settings were found by working the beam outward while making scans with the 5-finger radial probe\cite{Richardson1975,craddock1975properties}. One month of tuning was needed to reach 360\,MeV ($r=7$\,m). The cyclotron commissioning minutes of Dec.\ 6, 1974 state: {\it ``Optimum setting of trim coil power supplies (different from the April survey predictions) causing some concern.''}. Then on Dec.\ 13, the minutes state: {\it ``...have achieved 5\,nA to 273 inch [6.9\,m] radius ... To be done: Investigate beam loss in vicinity of 235 inch [6.0\,m] radius...''} 

\begin{figure}
        \centering
        \includegraphics[width=.75\textwidth]{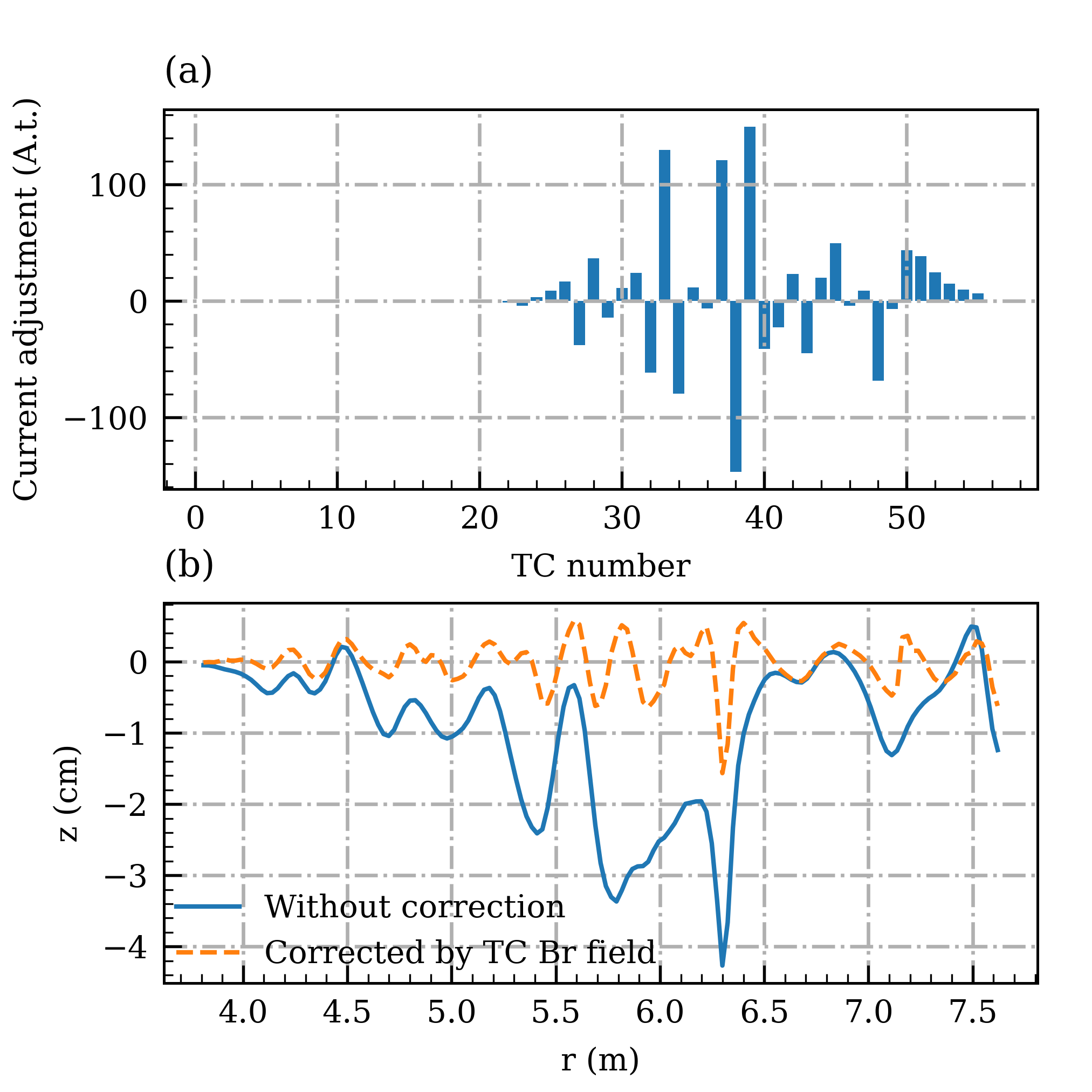}
        \caption{Correction of the closed orbit vertical position using trim coil $B_r$ field. (a) Adjustment of the trim coils'  settings. (b) Orbit vertical position before and after the correction.}\label{TCOP}
        \end{figure}

There is now an explanation for this. The ``CYC581'' {\tt CYCLOPS} calculation, based upon the field with the consistency error had the expected vertical orbit errors shown by the dashed orange curve at bottom in Fig.\,\ref{CYCLOPS}. But when we adjust the field for self-consistency as described above, it causes deviation from the median plane of up to 4\,cm (blue curve Fig.\,\ref{CYCLOPS}). But re-correction by adjusting the trim coils brings the beam back to within 1.5\,cm, as shown in Fig.\,\ref{TCOP}.

Unfortunately, since neither the raw base field without the trim coils' excitation nor the the original settings for the trim coils as used in the legacy field data files are known, there is no way to compare the new `trim-coil-corrected’ legacy field, with the cyclotron condition as it is running today. Nevertheless, the adjustment of the trim coil settings to achieve a $\pm$1.5 cm vertical centering under the corrected field map, which are all below 150 ampere-turns, do appear to be within a reasonable range of the operational settings.

%\subsection{Effect of the field survey data correction}
%We studied the tune and static equilibrium orbit of the cyclotron using the code CYCLOPS\cite{Gordon:1984EO, rao201350} that integrates the motion equations through the cyclotron field survey map shown in Fig.\,\ref{BZMAP} and the corrected median plane error field map. Fig.\,\ref{CYCLOPS} compares the results calculated using the CYC581 field survey map and the one using the corrected map. The corrected field map has the same $B_z$ and $B_r$ median plane map as CYC581 survey map, while the $B_\theta$ and $dBz/dz$ are corrected using the potential map calculated from $B_r$ median plane map. The difference between the corrected map and the survey map doesn't affect the cyclotron tune. However, it significantly changes the average vertical coordinates of the closed orbit.

%\subsection{Improving the beam vertical centering}
%To correct the large vertically off-centred orbit, we do some simulation studies to optimize the trim coil's $B_r$ field. TRIUMF cyclotron is equipped with 56 trim coils, each consisting of two circular loops, one above and one below the mid-plane, and spaced at 15\,cm radial intervals. They could be used with top and bottom coils in opposition (referred to as `$B_r$-mode'), to create a radial field and thereby center the beam vertically. Figure. \ref{TCOP} shows the simulation result of the vertical orbit correction. After correction, the vertical centering error is less than 1.5\,cm, which is small compared with the main magnet gap size of 48.26\,cm. At the same time, all the trim coil's current adjustment is within 150 ampere-turns, which is reasonable according to the capability of our trim coil power supplies.

\section{Conclusion}\label{sec:C}

By using Gordon's approach and a highly accurate differentiator, the redundant field survey maps $B_\theta$, $B_r$ and $dB_z/dz$, resulting from the magnetic median plane tilt error, have been crosschecked against each other. A systematic error in the $B_\theta$ survey data was found and has been corrected. The results of static orbit recalculation show that the correction doesn't affect the machine vertical tune, but does significantly increase the vertical centering errors of the orbits. This has provided an explanation as to why the originally-calculated trim coil settings were of little use in the original commissioning. Further study has shown that the increased centering errors can be reduced to below 1.5 cm by adjusting the trim coils' $B_r$ field within the output limits of the trim coils' power supplies. 
 
\section{Acknowledgments}

The author would extend sincere gratitude to Rick Baartman for investigating the TRIUMF cyclotron commissioning minutes, and Thomas Planche for fruitful discussions on the interpolation scheme.  This work was funded by TRIUMF which receives federal funding via a contribution agreement with the National Research Council of Canada.

\bibliographystyle{unsrt} 
\bibliography{./bib}
\end{document}